\documentclass[11pt]{article}
\usepackage{graphicx}
\usepackage{bm}        
\usepackage{amssymb}   
\usepackage{color}
\newcommand{\mpl}{M_{\rm Pl}}

\def\L*{{\cal L}_*}
\def\L{\mathcal{L}}
\renewcommand{\(}{\left(}
\renewcommand{\)}{\right)}

\def\nn{\nonumber}
\def\p{\partial}
\def\mn{_{\mu \nu}}
\def\stu{St\"uckelberg }
\def\p{\partial}

\def\<{\langle}
\def\>{\rangle}
\def\stu{St\"uckelberg }
\newcommand{\de}{\partial}
\newcommand{\be}{\begin{equation}}
\newcommand{\ee}{\end{equation}}

\newcommand{\tg}{\tilde{g}}
\newcommand{\al}{\alpha}
\newcommand{\bt}{\beta}
\newcommand{\gam}{\gamma}

\newcommand{\del}{\delta}

\newcommand{\eps}{\varepsilon}

\newcommand{\sig}{\sigma}

\newcommand{\om}{\omega}

\newcommand{\rmd}{\mathrm{d}}

\newcommand{\Mpl}{M_{\textrm{Pl}}}
\newcommand{\Mp}{M_{\textrm{Pl}}}

\renewcommand{\L}{\mathcal{L}}
\newcommand{\K}{\mathcal{K}}
\newcommand{\tK}{\tilde{\mathcal{K}}}
\newcommand{\x}{\vec{x}}

\usepackage{amsmath}
\usepackage{amsfonts}
\usepackage{verbatim}
\usepackage{setspace}
\onehalfspace
\begin{document}

\begin{titlepage}

\begin{flushright}
{NYU-TH-09/18/12 \\ UCSD-PTH-12-09}

\end{flushright}
\vskip 0.9cm

\centerline{\Large \bf Quasi-Dilaton: Theory and Cosmology}
\vskip 0.7cm
\centerline{\large Guido D'Amico$^{a}$, Gregory Gabadadze$^{a,b}$, Lam Hui$^c$, David Pirtskhalava$^d$} 
\vspace{0.1in}
\vskip 0.3cm

\centerline{\em $^a$Center for Cosmology and Particle Physics,
Department of Physics,}
\centerline{\em New York University, New York,
NY, 10003, USA}

\centerline{\em $^b$ Simons Center for Geometry and Physics,}
\centerline{\em Stony Brook University, Stony Brook,
NY, 11794, USA}

\centerline{\em $^c$Physics Department and Institute for Strings, Cosmology, and Astroparticle Physics,}
\centerline{\em Columbia University, New York, NY 10027, USA }

\centerline{\em $^d$Department of Physics, University of California, San Diego, La Jolla, CA 92093 USA}

\centerline{}
\centerline{}

\vskip 1.cm

\begin{abstract}

General Relativity (GR), with or without matter fields,  admits a natural  extension to 
a scale invariant theory that requires a dilaton. Here we show
that the recently formulated massive GR,  minimally coupled to matter, possesses 
a new global symmetry related to  scaling of the reference 
coordinates w.r.t. the physical ones.  The field enforcing this symmetry, 
dubbed here quasi-dilaton, coincides with an ordinary dilaton if only  pure gravity 
is considered,  but differs from it when the matter Lagrangian is present.  
We study: (1) Theoretical consistency of 
massive GR with the quasi-dilaton; (2) Consistency with observations for  
spherically symmetric sources on (nearly) flat backgrounds;  (3) Cosmological implications of this theory.
We find that: (I) The theory with the quasi-dilaton is as consistent as massive GR is. 
(II) The Vainshtein mechanism  is generically retained,  owing to the fact that in the decoupling limit 
there is an enhanced symmetry, which turns the quasi-dilaton  into a  second galileon,  consistently coupled to a tensor field.
(III) Unlike in massive GR, there exist flat FRW solutions. In particular, we find self-accelerated solutions and discuss their quadratic perturbations. These solutions are testable by virtue of the different effective Newton's constants that govern the Hubble expansion and structure growth.

\end{abstract}


\end{titlepage}

\section{Introduction and summary}

An extension of General Relativity (GR)  by a mass term, and more general polynomial terms,  
is motivated by the cosmological constant, and  dark energy problems. 
Such an extension was thought to be impossible due to the loss of the hamiltonian constraint,
leading to the existence of a ghost-like degree of freedom, 
in addition to the conventional 5 helicity states of a massive graviton \cite{Boulware:1973my}; 
this 6th degree of freedom is referred to as the Boulware-Deser (BD) ghost.

While the BD  work established the loss of the hamiltonian constraint  for 
a broad class of theories, later it was shown in Ref. 
\cite{Creminelli:2005qk}  that in {\it all} massive gravity theories the lapse would 
necessarily enter nonlinearly at the {\it quartic} order in fields, 
suggesting inevitability of the loss of the hamiltonian constraint, 
starting at the quartic order.
   
A loophole in these arguments  was found in Ref. \cite{deRham:2010ik}, by pointing out
that in a special  class of theories the hamiltonian 
constraint may  still be present even if the lapse enters the hamiltonian nonlinearly. 
Moreover,  Ref. \cite{deRham:2010ik}  gave an order-by-order 
Lagrangian free of the BD ghost in a particular limit, 
which prior to that was also thought to be impossible \cite{Creminelli:2005qk}.

Last but not least, in Ref. \cite{deRham:2010kj}  the order-by-order Lagrangian  of 
\cite{deRham:2010ik} was resummed into a diffeomorphism invariant nonlinear theory, 
which was proposed as a ghost-free candidate for massive GR.  
Moreover, it was  shown  in \cite{deRham:2010kj} that  the theory, in the unitary gauge,  
does exhibit the hamiltonian constraint in the quartic order,  even though the lapse
enters nonlinearly beginning from that order.

The absence of the BD problem in the unitary gauge to all orders
was established in Refs. \cite{Hassan:2011hr,Hassan:2011ea}, providing the proof of ghost-freedom
in the full theory. It looks like this  
can  also be  generalized  to a full covariant hamiltonian beyond  unitary gauge  
in which the  \stu fields  are retained  \cite{Hassan:2012qv}, \cite{Kluson:2012wf}.

In the Lagrangian formalism on the other hand, the absence of the BD ghost is related to a 
very special structure of the \stu sector \cite {deRham:2010kj, deRham:2011rn}.
Recently, Mirbabayi \cite{Mirbabayi:2011aa} has uncovered a number of remarkable features of this sector:
(1) He found  that there is an enhanced symmetry for quadratic  fluctuations  
on an {\it arbitrary} background, but in the leading order in the strength of the background.
(2) He showed  the  absence of  the BD ghost in small fluctuations 
on {\it any } background.  (3) Last, but not least, Mirbabayi  
showed that the absence of the BD ghost in the decoupling limit 
is not only necessary, but is also  sufficient for its absence in the full theory away from the 
decoupling limit. 

Since its formulation, and starting with Refs. 
\cite{deRham:2010tw} - \cite{Sbisa:2012zk}, massive GR has been 
used to study  cosmology, black holes, and other exact or approximate 
solutions (see Ref. \cite{Hinterbichler:2011tt} for a review of theoretical aspects of massive gravity and Ref. \cite{Goldhaber:2008xy} for a review on phenomenology of general massive gravity theories).
One interesting feature of cosmology in massive GR is the absence of spatially flat and closed FRW solutions, while the obtained inhomogeneous solutions may still well approximate the observed world \cite{D'Amico:2011jj}. There are also self-accelerated solutions for which the metric can be brought to the FRW form at an expense of having inhomogeneities in the \stu fields \cite{D'Amico:2011jj,Gratia:2012wt,Kobayashi:2012fz}.
Very interestingly however, a solution with FRW symmetry does exist for open  universe  \cite{Gumrukcuoglu:2011ew,Gumrukcuoglu:2011zh} (for subtleties on perturbations about these solutions, see \cite{DeFelice:2012mx,Guido}).
\vspace {0.1in}

Massive GR \cite {deRham:2010kj} has also been generalized to 
theories involving more fields. An arguably simplest generalization  is to introduce 
a scalar field the value of which sets the graviton mass \cite{D'Amico:2011jj}.  One may go further and 
introduce an additional dynamical tensor field to form bi-gravity a la Ref. 
\cite{Salam:1976as}, but now using the ghost-free theory 
of Ref. \cite{deRham:2010kj} as a basis of the construction, as was done in Ref. \cite{Hassan:2011zd}.
Such a bi-gravity, unlike the earlier versions \cite{Salam:1976as},   was  
shown  to be free of the BD ghost \cite{Hassan:2011zd}; its various classical 
solutions have been studied in \cite {Comelli:2011zm, Volkov:2011an}.  Furthermore, 
the tri-gravity generalization was also considered \cite{Khosravi:2011zi}.

Most interestingly, Hinterbichler and R. Rosen (HR) \cite {Hinterbichler:2012cn}, 
have recently shown that there is a certain overarching order in  the tensorial extensions: 
in $D$ space-time dimensions ($D\ge 3$) there are at most $D$ gravitons (one of them being massless) 
that can form consistent interacting  vertices of the tensor fields. HR proved this by 
reformulating the theory in a vielbein formalism, where they showed that the  
hamiltonian construction gets significantly simplified (see also references on earlier works 
using  the first order formalism in Ref. \cite {Hinterbichler:2012cn};  see 
Ref. \cite {Hassan:2012wt} for further development of the HR construction.).
As a bonus, by considering compactifications of the Hinterbichler-Rosen 
module in various dimensions one should be able to obtain consistent interacting  
theories of a finite number of tensors, vectors, and scalars\footnote{Owing to the fact that 
graviton  mass scale and the compactification scale need  not be 
related to each other, the latter can be taken to be much smaller 
than $1/m$, in which case all Kaluza-Klein modes could be decoupled.}.

\vspace{0.1in}

In the present work, we would like to discuss a particular 
extension of massive GR  by an additional scalar field. While certain scalar extensions have already 
been discussed by some of us  in  \cite{D'Amico:2011jj},  here we introduce  a special  
scalar $\sigma$ that gives rise to a certain new global symmetry of the Lagrangian.
The  symmetry transformation involves the scalar itself,  and the \stu 
fields $\phi^a,~a=0,1,2,3$, that are necessary if one wishes to work with a 
diffeomorphism-invariant action for massive GR.  
These four fields are scalars w.r.t. diffeomorphisms, but do transform under the 
Poincar\'e group of the internal space of $\phi^a$'s, as emphasized by Siegel 
in \cite{Siegel:1993sk}. The new global symmetry 
that we use as a building principle for the action involving
the scalar $\sigma$ is realized as follows: 
\be
\sigma \to \sigma - \alpha \, \Mpl \, , \quad \phi^a \to e^{\alpha}\phi^a \,,
\label{newglob}
\ee
where $\alpha$ is an arbitrary symmetry transformation parameter. 
The rest of the fields in the Einstein frame\footnote{We define the Einstein frame in the standard way - 
the one for which the kinetic term for the graviton has the usual Einstein-Hilber form;  Jordan 
frame on the other hand will feature a kinetic mixing between the scalar $\sigma$ and the graviton.},  
and the physical coordinates $x^\mu$,  do not transform.
This symmetry 
fixes uniquely, modulo some irrelevant derivative terms, an extension 
of massive GR by the $\sigma$ field; in particular, one consequence of \eqref{newglob} is minimal coupling of matter to gravity in the Einstein frame (unlike Brans-Dicke theories for instance, which have matter coupled minimally to gravity in the Jordan frame).

To motivate the symmetry (\ref {newglob}), we recall  that massive  gravity is built upon 
a reference Minkowski space with the metric
\be
g^{\rm ref}_{\mu\nu} = \frac{\partial \phi^a}{\partial x^\mu} \frac{\partial \phi^b}
{\partial x^\nu} \eta_{ab}\,,
\label{ref}
\ee
where $\eta_{ab} = {\rm diag} (-1,1,1,1)$ is the flat metric in the internal space of $\phi^a$'s.  
The fields $\phi^a$ can  be regarded  as arbitrary non-inertial 
coordinates of the reference Minkowski space\footnote{This does not mean, however, 
that the theory admits only Minkowski background;  nontrivial background solutions do certainly exist.};
the transformation of $\phi^a$'s in (\ref {newglob}) 
is then just a rescaling  of  the reference space coordinates. Thus, 
$\sigma$ is a field that enforces  invariance when  
the reference space coordinates get rescaled w.r.t. the physical space coordinates, 
while all the other fields  of the theory remain intact\footnote{Likewise,  
$\phi^a$'s  can  be regarded as target space coordinates, with  the world-volume coordinates  
being $x^\mu,~\mu=0,1,2,3$;   in this  case $\sigma$ enforces that 
the rescaling of the target space w.r.t. the world-volume be a symmetry.}.

Then, the following natural question arises:  what is the relation, if any,   
of the $\sigma$  field  to an ordinary dilaton?  

To answer this question, we temporarily transform our action  into the Jordan frame in Sec. \ref{themodel}.
Then, we find  that $\sigma$ indeed enters as an ordinary  dilaton would in the purely gravitational 
action.  Hence, the obtained action is also invariant w.r.t. conventional dilatations.  
In fact, the two global symmetries -- dilatations and new global symmetry  (\ref {newglob}) --  
are not independent, as far as the pure gravity 
action is concerned: the latter  is a linear combination of the former  and a global 
subgroup of diffeomorphisms. However, and perhaps not surprisingly, only one of these two global symmetries 
can be respected once the matter field  Lagrangian is introduced.  We choose this symmetry  
to be (\ref {newglob}), using it as a guiding principle for 
constructing our theory with the matter fields.  Thus, the obtained full 
theory is not invariant under dilatations, but preserves (\ref {newglob}).

This is easier to understand  by returning back to the Einstein frame: there, 
the matter fields are  coupled to the physical metric in a canonical way, with 
no direct  coupling  to  the \stu fields $\phi^a$. 
If so, the matter Lagrangian in the Einstein frame 
cannot  be  directly coupled to the $\sigma$ field in a nonderivative way, since 
this would violate the symmetry (\ref {newglob}).  This is not the case, however,  
for an ordinary dilaton, which does couple to the matter fields in the Einstein 
frame  without derivatives, thus violating (\ref {newglob}).  Therefore, 
our $\sigma$ is not a dilaton, nevertheless, it is closely related  
to the latter (is identical to it in pure gravity sector), so we will refer to it 
as ``quasi-dilaton'' in what follows (likewise, we will occassionally refer to \eqref{newglob} as ``quasi-dilatations'') .

Having established these properties, we move on to examine the theoretical 
consistency,  and then to deduce the physical 
consequences of massive gravity with the quasi-dilaton.

We start out by examining the absence of the BD ghost in the theory.
We  show that the quasi-dilaton 
turns into the second galileon in the decoupling limit (the first one being the helicity-0 mode 
of the massive graviton \cite {deRham:2010gu, deRham:2010ik}),  and as such it acquires an additional 
enhanced symmetry -- the field-space galilean invariance: 
\be
\sigma \to \sigma +c_\mu x^\mu + b\,,
\label{galsigma}
\ee
where $c_\mu$ and $b$ are arbitrary constants. As a result, the theory in the decoupling limit 
is free of the  BD ghost. We can therefore use Mirbababyi's method  to demonstrate that the full theory,  
away from the decoupling limit,  is also BD ghost-free. This is done in Sec. 3. 

Consistency of massive gravity with the quasi-dilaton would also be expected if the hamiltonian analysis of this theory 
were to be performed, since the quasi-dilaton does not introduce any derivatives in the mass 
term, and has a canonical kinetic term in the Einstein frame. Thus, we expect the constraint uncovered in \cite{Hassan:2011hr,Hassan:2011ea} to be only trivially modified in a theory with the quasi-dilaton.

As we have already mentioned, we introduce couplings  to matter 
in a minimal way so that only the tensor field in the Einstein 
frame couples to matter fields;  such a coupling does not 
spoil the consistency of the theory described above. 

As a next step we study phenomenological implications of the model, and in particular, 
whether the theory is capable of recovering the predictions of GR. To this end, we check the existence of 
the Vainshtein mechanism \cite{Vainshtein:1972sx, Deffayet:2001uk}.  Using the bi-galileon Lagrangian 
obtained in the decoupling limit, we show that static sources do exhibit the Vainshtein mechanism
-- the helicity-0 field is  suppressed at observable scales. Hence, the 
model should pass all observational tests of GR, at least for (nearly) flat backgrounds.

Furthermore, we discuss  cosmology of the theory.   We find  that unlike massive GR, it  admits 
flat FRW cosmology; in particular, selfaccelerated vacua exist
for a broad part of the parameter space.  We perform a preliminary examination of the fluctuations 
on these de Sitter (dS) solutions and  find that, unlike  
already existing examples in ghost-free massive GR, in general both of
the scalar degrees of freedom may propagate on such backgrounds.

The paper is organized as follows. In Sec. 2, we introduce the theory of quasi-dilaton massive gravity. 
Section 3 deals with the analysis of the theoretical consistency of the model, while the screening of extra scalar forces and cosmology are discussed in Sections 4 and 5 respectively. 

We conclude the present section by fixing notation and conventions used in the rest of the paper. 
As pointed out already, we use the mostly plus metric convention. The Levi-Civita symbol,
$\varepsilon_{\mu\nu\alpha\beta}$,  is normalized so that  $\varepsilon_{0123}=1$, while the symbol 
with upper indices is obtained by using the inverse metric tensor  $g^{\mu\nu}$ to rise the 
indices,  and by multiplying  the expression by ${\rm det}(g_{\mu\nu})$, so that  $\varepsilon^{0123}=-1$. 
Various  contractions of rank-2 tensors will often be 
denoted by square brackets as follows: $\K^\mu_{~\mu}=[\K]$, $\K^\mu_{~\nu}\K^\nu_{~\mu}=[\K^2]$, 
$\K^\mu_{~\alpha}\K^\alpha_{~\beta}\K^\beta_{~\mu}=[\K^3]$, etc. The (ordered) index contractions on the epsilon symbols  
will be omitted altogether; for example, 
$\varepsilon^{\mu\alpha\rho\sigma}\varepsilon^{\nu\beta}_{~~\rho\sigma}\Pi_{\mu\nu}\Pi_{\alpha\beta}
\equiv\varepsilon\varepsilon\Pi\Pi$, as well as 
$\varepsilon_\mu ^{~\gamma\alpha\rho}
\varepsilon_{\nu \gamma}^{~~\beta\sigma}\Pi_{\alpha\beta}\Pi_{\rho\sigma}\equiv\varepsilon_\mu\varepsilon_\nu
\Pi\Pi$, with an obvious generalization to terms with different number of $\Pi$'s.

\section{Quasi-dilaton massive gravity}
\label{themodel}
In this section we introduce the ghost-free model of quasi-dilaton massive gravity (QMG).
The theory is based on massive GR \cite{deRham:2010ik, deRham:2010kj}, representing a consistent nonlinear extension of the Fierz-Pauli theory \cite{Fierz:1939ix}.
We start out with a brief summary of these theories, followed by the construction of QMG, 
which can be obtained by supplementing the gravitational sector of massive GR with the global symmetry \eqref{newglob}.

\subsection{Massive GR}

A generic theory of a massive graviton can be 
written in a manifestly diffeomorphism-invariant way via an introduction of four spurious scalar fields, 
$\phi^a(x)$ \cite{Siegel:1993sk, ArkaniHamed:2002sp, Dubovsky:2004sg}.
One defines a covariant tensor $H_{\mn}$, related to the physical (coupled to matter) 
metric $g_{\mn}$ as follows:
\be
H_{\mn}=g_{\mn}-\p_\mu\phi^a\p_\nu\phi^b\eta_{ab} \, ,
\ee
where $a,b\in (0,1,2,3)$ are internal indices counting the scalars 
and $\eta_{ab}=\mathrm{diag}(-1,1,1,1)$ is the Minkowski metric in the internal space.
In the unitary gauge, the scalars are frozen to coincide with the corresponding 
coordinates $\phi^a(x)=\delta^a_\mu x^\mu$, however it is often more helpful not to resort to any particular gauge.

A covariant action for a consistent massive graviton with mass $m$  -- the one that propagates (at most) 5 degrees of freedom 
on a generic background, and in the presence of matter,  has the following specific form \cite {deRham:2010kj}:
\be
\label{eq:mg}
S = \frac{\Mp^2}{2} \int \rmd^4 x \sqrt{- g}  \left[ R
-\frac{m^2}{4} \( \mathcal{U}_2(\mathcal{K})+\alpha_3 \mathcal{U}_3(\mathcal{K})
+\alpha_4 \mathcal{U}_4(\mathcal{K}) \) \right]
+{S}_m(g,\psi),
\ee
where ${\cal K}$ is a four-by-four matrix  with the elements defined as 
\be
\K^\mu_{~\nu} =\delta^\mu_\nu-\sqrt{g^{\mu\alpha}\p_\alpha\phi^a \p_\nu\phi^b\eta_{ab}} \, ,
\label{K}
\ee
and $\mathcal{U}_i$ are specific polynomials of the matrix ${\cal K}$ 
\begin{subequations}
\begin{align}
\mathcal{U}_2&=4 ([\K^2]-[\K]^2)=2 \eps_{\mu \al . .} \eps^{\nu \bt . .} \K^\mu_{\; \nu} \K^\al_{\; \bt}\\
\mathcal{U}_3&=-[\K]^3+[\K][\K^2]-2[\K^3]=\eps_{\mu \al \gam .} 
\eps^{\nu \bt \del .} \K^\mu_{\; \nu} \K^\al_{\; \bt} \K^\gam_{\; \del} \\
\mathcal{U}_4&=-[\K]^4+6[\K ^2][\K]^2-[\K^3][\K]-3[\K^2]^2+6[\K^4]=
\eps_{\mu \al \gam \rho} \eps^{\nu \bt \del \sig} \K^\mu_{\; \nu} \K^\al_{\; \bt} \K^\gam_{\; \del} \K^\rho_{\; \sig} \,.
\label{potential}
\end{align}
\end{subequations}
Furthermore, $\alpha_i$'s are two arbitrary parameters characterizing a given theory and ${S}_m$ is the action for matter fields $\psi$, minimally coupled to the metric $g_{\mn}$. 
The dependence on the scalars $\phi^a$ comes in the graviton potential involving the $\K$ tensor \footnote{The 
square root of a matrix satisfies the property $\sqrt{A} \cdot \sqrt{A} = A$.
In general, there are 16 choices for the square root of a $4 \times 4$ matrix: 
the correct choice is the one with all four eigenvalues positive, as it is implied 
by the minus sign in the definition of $\mathcal{K}^\mu_{~\nu}$.}.

As shown in  \cite{deRham:2010ik},  the defining property of the 
potential terms $\sqrt{-g} \, \mathcal{U}_i$, which is also a necessary condition   
for the absence of the BD ghost in the decoupling limit around Minkowski space-time, 
is that upon substitution 
\be
\label{eq:sub}
h_{\mn}\equiv g_{\mn}-\eta_{\mn}=0, \qquad \phi^a(x)=\delta^a_\mu x^\mu-\eta^{a\mu}\p_\mu\pi,
\ee
all terms of the type $(\p^2\pi)^n$,  have to collect into a total derivative, thereby 
rendering the potentially dangerous higher-derivative self-interactions of $\pi$ nondynamical.
This eliminates the BD ghost from the theory in the DL, leading to a well-posed Cauchy problem 
and a single propagating degree of freedom in $\pi$, describing the helicity-0 graviton.
As shown by Mirbabayi, the above  necessary condition is also a sufficient one 
for the absence of the BD ghost in the full theory \cite{Mirbabayi:2011aa}. The polynomials 
$\mathcal{U}_i(\K)$ are specifically constructed to satisfy this property \cite {deRham:2010ik}: upon substitution \eqref{eq:sub}, the tensor $\K^\mu_{~\nu}$ reduces to $\p^\mu\p_\nu\pi$
and the antisymmetric structure of the potential makes it manifest that all $\pi$ self-interactions 
indeed collect into a total derivative. In four dimensions there are four independent such terms.
The zeroth order one is just a cosmological constant. The coefficient of the quadratic terms normalizes 
the graviton mass, leaving two free parameters in the theory. The Minkowski vacuum corresponds 
to $\phi^a=\delta^a_\mu x^\mu$, and the spectrum on flat space consists of the five polarizations 
of a massive graviton, while the potential specifies the consistent interactions of those.

\subsection{Adding the quasi-dilaton}

We would like to promote the purely gravitational sector of the ghost-free massive GR to a theory, 
invariant under the global rescalings of the four scalars $\phi^a$ w.r.t. the physical coordinates, 
$x^\mu$ (or, to put it in a different way, the dilatations in the internal space).
To this end, we introduce a canonically normalized field $\sigma$, and impose the global invariance, 
realized in the Einstein frame as in \eqref{newglob}. The extended  
massive gravity action which respects this symmetry reads,
\be
\begin{split}
\label{eq:einstein}
S_E =&  \int \rmd^4 x \frac{\Mp^2}{2} \sqrt{-g}  \left[ R -\frac{\omega}{\mpl^2} g^{\mn}\p_\mu\sigma\p_\nu\sigma
- \frac{m^2}{4} \( \mathcal{U}_2(\tK) +\alpha_3 \mathcal{U}_3(\tK)+\alpha_4 \mathcal{U}_4(\tK) \) \right] \\ 
&+ \int \rmd^4 x  \sqrt{- g} \mathcal{L}_m(g_{\mn},\psi) \, ,
\end{split}
\ee
where we have defined
\be
\tK^\mu_{~\nu}=\delta^\mu_\nu-e^{\sigma/\mpl}\sqrt{g^{\mu\alpha}\p_\alpha\phi^a
\p_\nu\phi^b\eta_{ab}} \, .
\ee
Note that the global symmetry \eqref{newglob} constrains the coupling of $\sigma$ to gravity up to derivative terms. 
We will choose this symmetry as a guiding principle for constructing matter couplings as well. In particular, 
we will couple the matter fields to the Einstein-frame metric $g_{\mn}$ in the minimal way, without any direct 
coupling to the \stu fields $\phi^a$. This significantly constrains the interactions of $\sigma$ with matter, 
allowing only for irrelevant derivative interactions, which we will ignore in what follows.

Note that the global symmetry \eqref{newglob} is a linear combination of the global subgroup of diffeomorphisms, 
$x^\mu\to e^{-\alpha} x^\mu$, and dilatations, which are also a symmetry of the purely 
gravitational sector and are realized \emph{in the Einstein frame} as
\be
\label{eq:scale}
x^\mu\to e^{\alpha}x^\mu \, ,
\qquad g_{\mn}\to e^{-2\alpha} g_{\mn} \, ,
\qquad \sigma \to \sigma - \mpl \alpha \, ,
\qquad \phi^a\to e^{\alpha}\phi^a \, .
\ee
For constructing the couplings to matter however, we choose to explicitly break the dilatation invariance \eqref{eq:scale}, retaining \eqref{newglob} as the exact global symmetry of the action \footnote{In fact, the choice of \eqref{newglob} as the guiding principle for coupling the theory to matter can easily be motivated from phenomenological considerations. Indeed, we cannot allow any $\mathcal{O}(1)$ coupling of matter to $\sigma$, if it is to be stabilized/hidden from the experimental tests of gravity. One can easily see this from the following reasoning. Massive gravity without the dilaton possesses a built-in property of screening extra scalars from observations - the Vainshtein mechanism \cite{Vainshtein:1972sx}, which originates from the continuity of the theory in the $m\to0$ limit, leading to the agreement of predictions of a massive theory with GR in the massless limit. But if $\sigma$ is to couple to matter, the massless limit of \eqref{eq:einstein} will feature a free dilaton, gravitationally coupled to external sources, modifying GR at $\mathcal{O}(1)$. In fact, even without coupling $\sigma$ to matter there still is a potential problem of hiding (one combination of) the scalars $\pi$ and $\sigma$ from solar system tests. We will however show in Sec. 4 that the Vainshtein mechanism successfully takes care of this issue.}.

One interpretation of the theory \eqref{eq:einstein} is the following.
In massive gravity, we introduce a fixed reference metric, which is usually chosen to be Minkowski.
In QMG, the quasi-dilaton appears in the action only through the combination 
$g^{\mn} \de_\mu \phi^a \de_\nu\phi^b \(e^{2 \sig/\mpl}  \eta_{ab}\) $, so we can think of it as the dynamical 
conformal mode of a field-dependent reference metric. This situation is reminiscent of bigravity 
theories \cite {Hassan:2011zd}, in which the entire reference metric is promoted to a dynamical field 
and the spectrum involves a massless and a massive gravitons. However, this analogy might be somewhat 
misleading: in bigravity the conformal 
mode would have a wrong sign kinetic term, whereas in QMG the quasi-dilaton has a right-sign  
kinetic term as long as $\om > 0$; hence,  the spectrum of the theory consists of a massive graviton 
and a massless scalar (see Appendix A)\footnote{The term $\int d^4 x ~e^{4\sigma/\mpl}\sqrt{\text{det} \p_\mu\phi^a\p_\nu\phi_a}$ is consistent with all the symmetries and requirements of the theory and can be added to the QMG action \eqref{eq:einstein} with an arbitrary coefficient (note that it is already present in \eqref{eq:einstein} as a part of $\mathcal{U} _4 (\tK)$). However, we will not be exploring this option here (in massive gravity without the quasi-dilaton, this term is trivial).}.



One can make a transition to the Jordan frame by performing the following conformal transformation
\be
g_{\mu \nu} = e^{2 \sig/\mpl} \tg_{\mu \nu} \, ,
\ee
under which the action becomes
\be
\begin{split}
\label{eq:jordan}
S _J = &\int \rmd^4 x \frac{\Mp^2}{2}  \sqrt{- \tilde g}  \bigg[ e^{2\sigma/\mpl}
\(\tilde R +\frac{(6-\omega)}{\mpl^2}\tilde{g}^{\mn}\p_\mu\sigma\p_\nu\sigma\) \\
&- \frac{m^2}{4}e^{4\sigma/\mpl} \( \mathcal{U}_2(\mathcal{K}) +\alpha_3 \mathcal{U}_3(\mathcal{K})+\alpha_4 \mathcal{U}_4(\mathcal{K}) \) \bigg]
+\sqrt{- \tilde g} e^{4 \sig/\mpl} \mathcal{L}_m(\tilde g_{\mn} e^{-2 \sig/\mpl}, \psi) \, .
\end{split}
\ee
Here $\tilde g_{\mn}$ denotes the Jordan frame metric with the corresponding notation for the curvature invariants constructed from it.

\section{Consistency of QMG}

In this section we study the consistency of QMG, which, as we show below, is closely related to the consistency of massive GR.
The latter theory was constructed to be ghost-free in the decoupling limit \cite {deRham:2010ik,deRham:2010kj}, 
and only later it was shown to be a consistent theory at the full non-linear level via Hamiltonian \cite{Hassan:2011hr}, as well as Lagrangian \cite{Mirbabayi:2011aa} analyses .
The current understanding is that the decoupling limit predictions in the given class of theories are rather powerful 
and already based on them one can make statements regarding the consistency away from the limit 
\cite{Mirbabayi:2011aa}.

With this in mind, we won't repeat the full canonical treatment of QMG for showing the presence of enough 
number of constraints, required for the absence of ghosts; the situation is rather similar to what happens 
in ghost-free massive GR, since the presence of the quasi-dilaton field causes minor modifications not affecting 
the conclusions regarding the absence of ghosts in the theory.
We will mostly be concerned with the decoupling limit analysis, which is best for displaying the physical content.
We also comment on the robustness of this  analysis for making predictions about 
the unitarity of the full theory  towards the end of this section.

\subsection{The decoupling limit and Galileons}

As mentioned above, we  can gain a better physical intuition about the degrees of freedom and nature of 
their interactions by working in the decoupling limit of QMG.
The limit corresponds to zooming to energy scales well above the mass of the graviton, while decoupling 
(at least as much as possible) gravity by sending $\mpl\to\infty$.
The degrees of freedom we will concentrate on are the quasi-dilaton, the helicity-2 polarization of the graviton $h_{\mn}$ and the (canonically normalized) helicity-0 graviton $\pi$, defined by $\phi^a(x)=\delta^a_\mu x^\mu-\eta^{a\mu}\p_\mu\pi/\mpl m^2$.
We will ignore the helicity-1 polarization $A_\mu$, since it is guaranteed to carry two degrees of freedom (due to an enhanced $U(1)$ symmetry in the decoupling limit) and guaranteed  not to couple to sources at the linear order. 
It  enters the decoupling limit  Lagrangian only quadratically, making $A_\mu=0$ a consistent solution to the 
equations of motion. The specific form of the potential $\sqrt{-g} \, \mathcal{U}({\cal K})$ guarantees that all 
dangerous self-interactions of the helicity-0 mode, being total derivatives, are rendered non-dynamical and drop 
out of the action, as discussed in the previous section. The decoupling limit is then defined as follows:
\be
\mpl\to\infty, \qquad m\to 0,\qquad  \Lambda_3=\(\mpl m^2\)^{1/3}=\mathrm{fixed} \, , 
\qquad \frac{T_{\mu \nu}}{\mpl} = \mathrm{fixed}\, .
\ee
In this limit, the Einstein frame Lagrangian \eqref{eq:einstein} in terms of the canonically normalized fields (with the metric perturbation normalized in the usual way, $h_{\mn}\to h_{\mn}/\mpl$), reduces to the following expression,
\be
\begin{split}
\label{eq:dl}
\mathcal{L}_{DL} =&
-\frac{1}{4} h^{\mn} \( \mathcal{E} h \)_{\mn}
-\frac{\omega}{2} \de^\mu \sigma \de_\mu\sigma \\
&- h^{\mu\nu} \left[ \frac{1}{4} \varepsilon_\mu \varepsilon_\nu \Pi
+ \( \frac{3}{16} \al_3 + \frac{1}{4} \) \frac{1}{\Lambda_3^3}\varepsilon_\mu \varepsilon_\nu \Pi \Pi
+ \( \frac{1}{16} \al_3 + \frac{1}{4} \al_4\) \frac{1}{\Lambda_3^6} \varepsilon_\mu \varepsilon_\nu \Pi \Pi \Pi \right] \\
&+ \sigma \left[ \frac{1}{2} \varepsilon \varepsilon \Pi
+ \( \frac{3}{8} \al_3-\frac{1}{2}\) \frac{1}{\Lambda^3_3} \varepsilon \varepsilon \Pi \Pi
+  \(\frac{1}{2} \al_4 - \frac{3}{8} \al_3\)\frac{1}{\Lambda^6_3}\varepsilon \varepsilon \Pi \Pi \Pi
-  \frac{\al_4}{2} \frac{1}{\Lambda^9_3}\varepsilon \varepsilon \Pi \Pi \Pi \Pi \right] \\
&+\frac{1}{\mpl} h^{\mn} T_{\mn}(\psi) \, . 
\end{split}
\ee
Here $\Pi_{\mn}=\p_\mu\p_\nu\pi$ is the rank-2 tensor constructed from the second derivatives of the helicity-0 graviton, $T_{\mn}$ is the matter stress-energy tensor, while
\be
\label{eq:Eterm}
(\mathcal{E} h)_{\mn}=- \frac{1}{2} \bigg( \square h_{\mu \nu} - \partial_\mu \partial^\alpha
h_{\alpha\nu} - \partial_\nu \partial^\alpha
h_{\alpha\mu} + \partial_\mu \partial_\nu h -
\eta_{\mu\nu} \square h + \eta_{\mu\nu}
\partial^\alpha  \partial^\beta h_{\alpha\beta}\bigg) \, 
\ee
is the linearized Einstein tensor on the Minkowski background.

The decoupling limit theory, despite the presence of higher derivatives in the action, is consistent and does not propagate any additional ghost-like degrees of freedom. The interactions between the helicity-2 and scalar gravitons come from the potential term in the action and, being certain scalar-tensor generalization of Galileons, have been shown to be ghost-free in \cite{deRham:2010gu} (in the antisymmetric form used above, it is obvious that the presence of the Levi-Civita symbol 
forbids more than two time derivatives on any field in the equations of motion). Quite remarkably, the sector of the theory involving the interactions of the quasi-dilaton with the scalar graviton is nothing but a bi-Galileon theory, 
rendering the decoupling limit of QMG completely free of ghosts. 

\subsection{Comments on the consistency of the full theory}

The unitarity of QMG in the decoupling limit, as the experience with the ghost-free massive GR indicates, can be considered as a strong hint of the absence of the BD ghost in the full nonlinear theory.
The canonical way of obtaining the number of propagating degrees of freedom is resorting to the Hamiltonian analysis.
Fortunately, the ghost-free massive GR has been extensively studied in the Hamiltonian formulation \cite{Hassan:2011hr} and the presence of just enough number of constraints for the propagation of exactly five d.o.f.'s of a massive spin-2 particle has been shown. It is straightforward to see that the extra quasi-dilaton present in the theory can not spoil these constraints, leading to the six propagating degrees of freedom in QMG.

In fact, as shown in \cite{Mirbabayi:2011aa}, already the special form of the decoupling limit \eqref{eq:dl} is sufficient for proving the absence of the BD ghost in the full theory.
At the quadratic order, QMG propagates a massive spin-2 state plus a scalar $\sigma$ (see Appendix A) and the special Fierz-Pauli structure eliminates the potential BD ghost, which might nevertheless show up at the nonlinear order in the pathological interactions of the helicity-0 graviton either with itself or the rest of the fields; on generic backgrounds on the other hand, if present, the BD ghost will reappear at the quadratic order giving a ghost pole to the graviton propagator.
The idea is then to study the theory around an arbitrary background in a locally inertial frame.
Showing the unitarity of the quadratic theory in this setting can therefore be considered the proof of the absence of the ghost in the full theory.
Let us briefly review the argument.
The form of the decoupling limit Lagrangian \eqref{eq:dl} significantly constrains what the full theory can look like.
Based solely on this limit, one can argue that the part of the full action involving the "vector" $A_{\mu}$ describing the deviation of the four scalars from the unitary gauge $\phi^a=\delta^a_\mu x^\mu-\eta^{a\mu}A_{\mu}$, has the following schematic form 
\be
\begin{split}
\label{eq:dofs}
\mathcal{L} &\supset h^{\mn}\(a_1~\varepsilon_\mu\varepsilon_\nu S+a_2~\varepsilon_\mu\varepsilon_\nu S S+ a_3~\varepsilon_\mu\varepsilon_\nu S S S\) \\
&+h F F (b_1+b_2~ \p A+\dots) +
\sigma\(c_1~\varepsilon\varepsilon S+c_2~\varepsilon\varepsilon S S+ c_3~\varepsilon\varepsilon S S S+c_4~\varepsilon\varepsilon S S S S\) \\
&+\sigma F F (d_1+d_2~ \p A+\dots)
 -\frac{1}{4}F F\(e_1+e_2~ \p A+\dots\) +\sum_{m,n}h^n\sigma^m+ \dots
\end{split}
\ee
Here $\(a_i,b_i,c_i,d_i,e_i\)$ are some constants, $\(m,n\)$ denote arbitrary (semi)positive integers, we have defined $S_{\mn}=\p_{( \mu} A_{\nu )}$, $F_{\mn}=\p_{[ \mu} A_{\nu ]}$, and ellipses denote the GR interactions along with the $\sigma$ kinetic term.
The key insight of \cite{Mirbabayi:2011aa}, based on the massive GR action without the quasi-dilaton, is that on any background characterized by VEVs of the fields appearing in \eqref{eq:dofs}, the specific form of the interactions guarantees that $h_{00}$ and $h_{0i}$ perturbations represent Lagrange multipliers in the quadratic lagrangian, while one combination of $A_\mu$-perturbations is non-dynamical, leaving five propagating degrees of freedom.
One can see from \eqref{eq:dofs} that the interactions of $\sigma$ with $A_{\mu}$ are just the ``scalar versions" of those of $h_{\mn}$, characterized by the same antisymmetric structure. Not surprisingly therefore, one can straightforwardly see that exactly the same conclusions apply to QMG, leading to (at most) six degrees of freedom  (the massive spin-2 field and the quasi-dilaton) around an arbitrary background - and therefore the absence of the BD ghost in the full theory.

\section{Screening of Extra Forces}

The presence of extra scalar degrees of freedom in QMG is potentially dangerous for phenomenology, since gravitationally coupled scalars could lead to an unobserved long-range fifth force.
The same problem is at first sight present in QMG, since $\sigma$ is a massless field which mixes with the helicity-0 graviton and one combination of these two fields will couple to matter; however, as we show in this section, this model possesses a built-in mechanism for hiding the extra scalars it propagates.
The indication of such screening, called the Vainshtein mechanism \cite{Vainshtein:1972sx}, can already be read off the decoupling-limit lagrangian featuring the Galileon interactions.
Massive GR without the dilaton has been shown to possess the  Vainshtein screening for a vast part of the parameter space \cite{deRham:2010ik,Koyama:2011xz,Chkareuli:2011te}, and so have the bi-Galileon theories \cite{Padilla:2010tj}.
Working in the decoupling limit of QMG which represents a certain combination of these two theories, we show that the screening mechanism successfully operates in this model as well; around localized sources both the quasi-dilaton and the helicity-0 graviton profiles are significantly suppressed within a certain distance $r_*$, the Vainshtein radius\footnote{Which e.g. for ordinary sources like the Sun is much larger than the galactic scales.}, and the gravitational potential is completely determined by the helicity-2 contribution, recovering GR in this region to a very high precision\footnote{We stress again that the helicity-1 mode of the massive graviton is irrelevant in this context since it does not couple to the external stress tensor at the linear level.}. 
We will concentrate on the decoupling limit analysis, since it completely captures all essential aspects of the Vainshtein mechanism.
This limit is a valid description of physics at distance scales $\Lambda_3^{-1}\ll r \ll m^{-1}$, where the lower limit marks the regime in which quantum corrections become important in the effective theory, while the upper one is due to the definition of the DL.
Moreover, we will make a particular choice of the parameters,
\be 
\alpha_3=-4\alpha_4, \label{param}
\ee
for which the interactions between the helicity-2 and helicity-0 gravitons can be completely eliminated by a redefinition of the helicity-2 field \cite{deRham:2010ik}; as a result, the theory breaks into separate helicity-2 and scalar sectors, the former characterizing the GR part of the gravitational potential.
To this end, we make the following redefinition of the helicity-2 graviton in the decoupling limit theory \eqref{eq:dl},
\be
\label{eq:hredef}
h_{\mn}\to h_{\mn}+\pi\eta_{\mn}+\frac{\gam_{-}}{\Lambda^3_3}\pi\Pi_{\mn},
\ee
under which the part of the lagrangian involving $h_{\mn}$ reduces to the usual linearized GR coupled to the external stress tensor, while the scalar sector lagrangian is given as follows:
\be
\begin{split}
\label{eq:scalarlagr}
&\mathcal{L}_{s} = -\frac{1}{8}\pi \( \varepsilon\varepsilon\Pi + 2 \frac{\gam_{-}}{\Lambda^3_3} \varepsilon\varepsilon\Pi\Pi
+ \frac{\gam_{-}^2}{\Lambda^6_3} \varepsilon\varepsilon \Pi\Pi\Pi \)
-\frac{\omega}{12}\sigma\varepsilon\varepsilon\Sigma \\
&+\frac{1}{2} \sigma \( \varepsilon\varepsilon\Pi - \frac{\gam_{+}}{\Lambda^3_3} \varepsilon\varepsilon\Pi\Pi
+4 \frac{\alpha_4}{\Lambda^6_3}  \varepsilon\varepsilon\Pi\Pi\Pi
- \frac{\alpha_4}{\Lambda^9_3} \varepsilon\varepsilon\Pi\Pi\Pi\Pi \)
+\frac{1}{\mpl}\pi T +\frac{\gam_{-}}{\mpl\Lambda^3_3}\pi\Pi_{\mn}T^{\mn}. 
\end{split}
\ee
Here we have defined $\gam_{\pm}=1 \pm 3\alpha_4$; also, $\Sigma_{\mn}\equiv\p_\mu\p_\nu\sigma$ and $T\equiv\eta^{\mn} T_{\mn}$ denotes the trace of the matter stress-tensor.
The above action involves kinetic mixing of the scalars, however it will prove to be simpler to study the classical solutions in this non-canonically normalized form. 

The equations of motion for $\sigma$ and $\pi$ respectively are
\begin{align}
&-\frac{\omega}{6}\varepsilon\varepsilon\Sigma
+\frac{1}{2}\varepsilon\varepsilon\Pi-\frac{\gam_{+}}{2\Lambda^3_3}\varepsilon\varepsilon\Pi\Pi
+2\frac{\alpha_4}{\Lambda^6_3}\varepsilon\varepsilon\Pi\Pi\Pi-\frac{\alpha_4}{2\Lambda^9_3}\varepsilon\varepsilon\Pi\Pi\Pi\Pi = 0 \, , \\
&-\frac{1}{4}\varepsilon\varepsilon\Pi-\frac{3\gam_{-}}{4\Lambda^3_3}\varepsilon\varepsilon\Pi\Pi
-\frac{\gam_{-}^2}{2\Lambda^6_3}\varepsilon\varepsilon\Pi\Pi\Pi+\frac{1}{2}\varepsilon\varepsilon\Sigma
-\frac{\gam_{+}}{\Lambda^3_3}\varepsilon\varepsilon\Sigma\Pi \nonumber \\
&\qquad \qquad \qquad \qquad \qquad 
+ 6\frac{\alpha_4}{\Lambda^6_3}\varepsilon\varepsilon\Sigma\Pi\Pi
-\frac{2\alpha_4}{\Lambda^9_3}\varepsilon\varepsilon\Sigma\Pi\Pi\Pi=-\frac{1}{\mpl} T \, .
\end{align}
For localized spherically symmetric, static configurations for which the only nonzero component of the stress tensor is $T_{00}=\bar{ M}\delta(r)/r^2$, these equations can be integrated once, which, in the spirit of Galileon theories, reduces them to algebraic equations for the first radial derivatives of fields
\begin{align}
&\omega \lambda_\sigma-3 \lambda_\pi+2 \gam_{+}\lambda_\pi^2-4\alpha_4\lambda_\pi^3=0 \, , \\
& \frac{3}{2}\lambda_\pi+3 \gam_{-}\lambda^2_\pi+\gam_{-}^2\lambda_\pi^3-3\lambda_\sigma+4 \gam_{+}\lambda_\sigma\lambda_\pi-12\alpha_4\lambda_\pi^2\lambda_\sigma=\(\frac{r_*}{r}\)^3 \, ,
\end{align}
where we have used the following notation:
$$
r_*=\(\frac{\bar M}{\mpl^2 m ^2}\)^{1/3},\qquad \lambda_\pi=\frac{\pi'}{\Lambda_3^3 r}, \qquad \lambda_\sigma=\frac{\sigma'}{\Lambda_3^3 r}~.
$$
Note that, in this basis, only $\pi$ is coupled to sources and the correction to the gravitational potential is entirely determined by its profile. One can express $\lambda_\sigma$ in terms of $\lambda_\pi$ from the $\sigma$ - equation, which reduces the system to its final form\footnote{It is worth to stress once again that we keep here the notation in terms of  $\gam_{\pm}$ not to overload the expressions; one should however keep in mind that these constants really depend on $\alpha_4$.},
\be
\label{v1}
\lambda_\sigma=\frac{1}{\omega}\( 4\alpha_4\lambda^3_\pi-2\gam_{+}\lambda_\pi ^2+3\lambda_\pi \) \, ,
\ee
\begin{multline}
\label{v2}
\frac{3}{2}\(1-\frac{6}{\omega}\)\lambda_\pi +\(3 \gam_{-}+\frac{18 \gam_{+}}{\omega}\)\lambda_\pi ^2+\bigg(\gam_{-}^2-\frac{8 \gam_{+}^2+48\alpha_4}{\omega}  \bigg)\lambda_\pi^3 \\
+\frac{40 \gam_{+}\alpha_4}{\omega}\lambda_\pi^4-\frac{48\alpha_4^2}{\omega}\lambda_\pi^5
=\(\frac{r_*}{r}\)^3 \, .
\end{multline}

There are a number of observations one can make about the last equation.
In the limit of large distances, $r\gg r_*$, the solution is obtained by simply neglecting all nonlinearities, $\lambda_\pi\sim \(r_*/r\)^{3}$ and $\pi$ has the usual Newtonian profile, modifying the gravitational force at $\mathcal{O}(1)$ beyond the Vainshtein radius - a manifestation of the famous vDVZ discontinuity \cite{vanDam:1970vg, Zakharov:1970cc}.
Moreover, from \eqref{v2} one can see that $\omega>6$ is required for the extra scalar force to be attractive in the vDVZ region.
Not surprisingly, this coincides with the condition of absence of a scalar ghost on Minkowski vacuum, which can be obtained by diagonalizing the decoupling limit scalar action \eqref{eq:scalarlagr}.
On the other hand, well within the Vainshtein radius $r\ll r_*$, the largest nonlinearity on the l.h.s. (assuming $\omega\sim 1$) dominates and the solution becomes
\be
\lambda_\pi\simeq -\(\frac{\omega}{48 \alpha_4^2}\)^{1/5} \(\frac{r_*}{r}\)^{3/5}\Rightarrow \pi\simeq -\frac{5}{7}\(\frac{\omega r_*^3}{48 \alpha_4^2}\)^{1/5}\Lambda_3^3 r^{7/5} \, ,
\label{pi}
\ee
while the correction to the gravitational potential is tiny\footnote{Although the profile of $\sigma$ is not probed by matter, it is interesting to observe that it is also screened by the Vainshtein mechanism,
\be
\sigma\sim -\frac{\Lambda^3_3 r_*^{9/5}}{\omega^{2/5}} r^{1/5},
\ee 
being significantly suppressed with respect to the Newtonian potential within $r_*$.
}
\be
\frac{\pi}{h_{00}}\sim \(\frac{r}{r_*}\)^{12/5} \, .
\ee
In fact, this screening is parametrically larger than in the DGP model \cite{Lue:2002sw, Dvali:2002vf,Gabadadze:2005qy}, leading to the impossibility of observing these corrections in the near future. Note that the seeming enhancement for $\omega\gg 1$ of the correction to the Newtonian potential in \eqref{pi} might be misleading; for large values of $\omega$ the quasi-dilaton becomes decoupled and the predictions of the theory should reproduce those of the ghost-free massive GR, for which the choice of the parameters given in  \eqref{param} leads to the scalar dynamics governed by quartic Galileon self-interactions. This can be directly seen
in \eqref{v2}, where, for $\omega\to\infty$, the leading nonlinearity is precisely that coming from the quartic Galileon, leading to the corresponding screening for $\pi$.

We have therefore shown that within the Vainshtein radius of a point source the gravitational force reduces to an excellent accuracy to the usual Newtonian attraction mediated by the helicity-2 polarizations of the graviton.
Vainshtein screening is a robust phenomenon and can be expected to work for a generic class of sources.
Also, here we have considered only a part of the parameter space, but we expect this mechanism to be operative for a more general choice of parameters in the theory.
We are not concerned with a detailed analysis of the solutions at this stage, however what we have shown above can already be considered a direct indication of the presence of a built-in mechanism for screening extra scalars in QMG.

\section{Cosmology}

Another important property of QMG is the existence of homogeneous and isotropic flat FRW solutions. In ghost-free massive GR, the same Hamiltonian constraint that removes the dangerous BD ghost from the theory leads to overrestrictive cosmological implications. Namely, it constrains the FRW scale factor to be independent of time, leaving Minkowski space as the only flat homogeneous and isotropic cosmological solution \cite{D'Amico:2011jj}. However, inhomogeneous/anisotropic solutions can arbitrarily closely approximate the standard FRW cosmology for sufficiently small graviton mass, due to the continuity in the $m\to 0$ limit \cite{D'Amico:2011jj}. The improvement of the situation in QMG is achieved at an expense of the presence of the quasi-dilaton in the theory; then, instead of constraining the scale factor to be a constant, the extra constraint simply relates its time evolution to that of the quasi-dilaton, implying the possibility for nontrivial flat FRW solutions. However, the following natural question arises: for $\omega\to\infty$, the quasi-dilaton should decouple from the rest of the fields, leaving the dynamics dominated by massive GR; how is then the $\omega\to \infty$ limit of the homogeneous and isotropic FRW universe of QMG expected to reproduce the inhomogeneous cosmology of the former theory? The obvious clue to this seeming puzzle is that there should be a certain kind of obstruction to QMG's flat FRW cosmology for the values of $\omega$, greater than some critial value. We will explicitly see how this works below. 
\subsection{Cosmological solutions and self-acceleration}

In the ghost-free massive GR without the quasi-dilaton, as mentioned above, the extra constraint that removes the BD ghost also forces the trivial Minkowski space to be the only homogeneous and isotropic flat FRW solution. In QMG however, the presence of the extra scalar lifts this constraint, giving it a new r\^ole of relating the time evolution of the scale factor to its own.
In order to see how this works, one can concentrate on the homogeneous and isotropic field configurations in the Einstein frame theory \eqref{eq:einstein}.
We start from the most general ansatz for flat solutions with this symmetry:
\be
\rmd s^2=- N^2(t) \rmd t^2+a^2(t) \rmd \x^2 \, , \quad \phi ^0 = f(t) \, , \quad  \phi ^i = x^i \, , \quad \sig = \sig(t) ,
\ee
and substitute this into \eqref{eq:einstein} to obtain the \emph{minisuperspace} action\footnote{Note that we have retained the lapse $N(t)$ in the action despite the fact that by time reparametrization invariance it can be fixed to an arbitrary value as long as $f(t)$ does not equal to one. However, keeping it explicitly is quite convenient, since it allows to quickly derive a first-order Friedmann equation for the scale factor.}
\be
\begin{split}
S =& \frac{\Mpl^2}{2} \int \rmd^4 x ~ \bigg\{- 6 \frac{a}{N} \(\frac{da}{dt}\)^2+ \frac{\omega}{\Mpl^2} \frac{a^3}{N} \(\frac{d\sigma}{dt}\)^2 \\
&+ 6 m^2 \bigg[ (2+ \al_3 + \al_4) a^3 N 
- \(1 + \frac{3}{4} \al_3 + \al_4\) e^{ \sig/\Mp} \(3 a^2 N + a^3 \frac{df}{dt}\) \\
&+ \(1+\frac{3}{2} \al_3 + 3 \al_4\) e^{2  \sig/\Mp} \(a N + a^2 \frac{df}{dt}\)
- \frac{1}{4} \(\al_3 + 4 \al_4\) e^{3  \sig/Mp} \(N+3 a \frac{df}{dt}\) \\
&+ \al_4 e^{4  \sig/\Mp} \frac{df}{dt}\bigg] \bigg \}
+ S_m.
\end{split}
\ee
In what follows we assume the matter sector $S_m$ to consist of a perfect fluid with the energy density $\rho_m$ and pressure $p_m$.
There are four fields in the above Lagrangian, however time reparametrization invariance
$$
t\to g(t),  \qquad N\to\frac{N}{\dot g(t)}~,
$$
guarantees that the $a$-equation of motion is redundant, being a certain linear combination of the rest.
Variation w.r.t. $f$ gives the constraint equation
\begin{multline}
\label{eq:phi}
\(1 + \frac{3}{4} \al_3 + \al_4\) a^3 e^{\sig/\Mp} - \(1+\frac{3}{2} \al_3 + 3 \al_4\) a^2 e^{2  \sig/\Mp} \\
+ \frac{3}{4} \(\al_3 + 4 \al_4\) a e^{3  \sig/Mp} - \al_4 e^{4 \sig/\Mp} = k \, ,
\end{multline}
where $k$ is an integration constant.
The Friedmann equation can be obtained by varying w.r.t. $N$,
\be
\begin{split}
&3 \Mp^2 H^2 + 3 \Mp^2 m^2 \bigg[ (2 + \al_3 + \al_4)
- (3 + \frac{9}{4} \al_3 + 3 \al_4) \frac{e^{ \sig/\Mp}}{a} \\
&+ (1+\frac{3}{2} \al_3 + 3 \al_4) \frac{e^{2  \sig/\Mp}}{a^2}
- \frac{1}{4} (\al_3 + 4 \al_4) \frac{e^{3  \sig/Mp}}{a^3} \bigg] = \frac{\omega}{2} \dot{\sig}^2 + \rho_m \, ,
\end{split}
\ee
here and in what follows the dot denotes a derivative with respect to cosmic time, i.e. $\dot{\sig} \equiv d \sig/(N d t)$ and consequently $H \equiv \dot{a}/a = d \ln a/(N d t)$.
Notice that eq. \eqref{eq:phi} is a fourth order algebraic equation for a generic choice of parameters.
However a very simple solution exists if $k = 0$, in which case the following ansatz can be used
\be
\label{eq:const}
e^{\sig/\Mp} = c a(t)~,\qquad \frac{d\sigma}{dt}=\mpl H,
\ee
with $c$ a constant to be determined below.
For this ansatz, the Friedmann equation reads
\be
\label{eq:frwc}
\begin{split}
\(3 - \frac{\omega}{2} \) \Mp^2 H^2 =& 
~3 \Mp^2 m^2 \bigg[ \frac{1}{4} (\al_3 + 4 \al_4) c^3 
- (1+\frac{3}{2} \al_3 + 3 \al_4) c^2 \\
&+ (3 + \frac{9}{4} \al_3 + 3 \al_4) c
- (2 + \al_3 + \al_4) \bigg] + \rho_m \, , 
\end{split}
\ee
while $c$ itself can be obtained from the constraint equation \eqref{eq:phi} with $k=0$,
\be
\label{eq:phi''}
c \left[ \(1 + \frac{3}{4} \al_3 + \al_4\) - \(1+\frac{3}{2} \al_3 + 3 \al_4\) c
+ \frac{3}{4} (\al_3 + 4 \al_4) c^2 - \al_4 c^3 \right] = 0 \, .
\ee
Before analyzing these equations more closely, it is timely to make an observation about the Friedmann equation \eqref{eq:frwc}: its left hand side flips the sign unless $\omega<6$. This sign flip is unacceptable if the theory is to describe real cosmology in the matter dominated era in which case the right hand side is strictly positive. Therefore, quite interestingly, we are led to conclude that the parameter space leading to self-accelerated vacua capable of describing the universe is orthogonal to the parameter space for which the Minkowski vacuum is ghost-free. This is also consistent with the expectation that for $\omega\to\infty$, the sigma field decouples and one shouldn't expect to have a homogeneous and isotroic solution.

One obvious solution to \eqref{eq:phi''} is $c=0\Rightarrow \sigma \to -\infty$. Substituting this value for the quasi-dilaton background into the Friedmann equation, one obtains the conventional FRW cosmology with an effective cosmological constant (energy density)
\be
\Lambda_{eff}=-3\mpl^2m^2\(2+\alpha_3+\alpha_4\),
\ee
which is positive for the choice of parameters, such that 
\be
\label{cond1}
\alpha_3+\alpha_4<-2.
\ee
A quick inspection of the Einstein frame action \eqref{eq:einstein} reveals that at least at the level of the background this is a well-defined solution, since (except for the quadratic kinetic term) $\sigma$ appears only through $e^{+\sigma /\mpl}$ everywhere in the Lagrangian. One should however be cautious about the zero expectation value of the latter quantity, since it multiplies the entire action involving the perturbations of the auxiliary scalars, pointing towards an infinitely strongly coupled vector/scalar graviton sector at the level of perturbations. We will not pursue this solution further in this paper.

Another solution is given by
\be
c = 1 \, ,
\ee
which leads to a vanishing cosmological constant in eq. \eqref{eq:frwc}. One can show from the $\sigma$-equation of motion (see below) that on this solution $a\dot f$ asymptotically approaches unity, leading to a Lorentz-invariant Minkowski background at late times\footnote{A remark on physical equivalence of different solutions in the theory is in order here. The presence of the global symmetry, the quasi-dilatations \eqref{newglob}, allows us to shuffle any overall normalization between the quasi-dilaton $\sigma$ and the auxiliary scalars $\phi^a$. The real invariant, both under the diffeomorphisms and the quasi-dilatations, is $e^{2\sigma/\mpl}g^{\mn}\p_\mu\phi^a\p_\nu\phi^b\eta_{ab}$, leading to physical equivalence of spatially homogeneous solutions with the same value for this invarant. It is therefore obvious, that the asymptotics of the solution at hand, $e^{\sigma/\mpl}=a$, $ds^2=-dt^2+a^2 \rmd \x^2$, $\phi^0=t/a$, $\phi^i=x^i$ is physically equivalent to the usual Minkowski vacuum of the theory.}. However, as we have noted above, the absence of a scalar ghost requires $\omega>6$ on this background, whereas the parameter space leading to a consistent cosmology is exactly orthogonal, $\omega<6$. Therefore, the solution for which $c=1$ is unacceptable from the consistency point of view and we will discard it in what follows.

The remaining two solutions of the constraint equation are given by the following expressions:
\be
\label{c's}
c_{2,3} = \frac{3 \al_3 + 8 \al_4 \pm \sqrt{9 \al_3^2 - 64 \al_4}}{8 \al_4} ,
\ee
which, when plugged back into the Friedmann equation \eqref{eq:frwc}, lead to an effective cosmological constant, describing either de Sitter or anti-de Sitter spaces in the absence of matter. It is straightforward to analyze these solutions for finding the parameter space with the \textit{physical} de Sitter vacua (i.e. the ones for which both the left and right hand sides of the Friedmann equation are positive). The result is
\be
\label{cond2}
\alpha_3\neq 0, \qquad 0<\alpha_4<\frac{\alpha^2_3}{8}.
\ee
We therefore conclude that for the vast region of the parameter space given by the conditions \eqref{cond2}, there exist de Sitter solutions, capable of describing the late-time acceleration of the universe.

For completeness, we also need to analyze the equation for $\dot{f}$.
Varying the action w.r.t. $\sig$, we obtain
\begin{multline}
\label{eq:sigma}
\frac{\om}{a^3} \frac{~\ d}{N dt}(a^3 \dot{\sig}) + 3 \Mpl m^2  c \bigg[ 
\(1 + \frac{3}{4} \al_3 + \al_4\) (3 + a \dot{f})
- 2 \(1+\frac{3}{2} \al_3 + 3 \al_4\) (1 + a \dot{f}) c \\
+ \frac{3}{4} \(\al_3 + 4 \al_4\) (1+3 a \dot{f}) c^2
- 4 \al_4 a \dot{f} c^3 \bigg] = 0 \, ,
\end{multline}
which gives $\dot{f}$ in terms of the scale factor.
Using the constraint equation \eqref{eq:phi''}, this simplifies to
\be
\label{adotf}
a \dot{f} = 1 + \frac{\om}{3\kappa m^2}~ (3 H^2 + \dot{H}) \, ,
\ee
where
\be
\label{eq:kappa}
\kappa = c ~\bigg [ 3 \(1 + \frac{3}{4} \al_3 + \al_4\)
- 2 \(1+\frac{3}{2} \al_3 + 3 \al_4\) c \\
+  \frac{3}{4} \(\al_3 + 4 \al_4\) c^2 \bigg ] \, .
\ee
As already noted above, the equation \eqref{adotf} shows that in the absence of a cosmological constant, the solution approaches the Lorentz-invariant Minkowski background at late times, which is unstable for the choice of parameters that can lead to realistic cosmological backgrounds.
We do not have such a problem, however, if the late-time solution is cosmological constant dominated; this means that the most natural endpoint of the cosmological evolution in the theories at hand is one of the above-described self-accelerated de Sitter backgrounds with tiny curvature $\sim m^2$.

\subsection{Perturbations}

Here we give some important, preliminary results on perturbations over de Sitter vacua, obtained above. To summarize the results on the background evolution, in the (cosmic) coordinates used in the previous subsection, the dS solution is (for simplicity, we will set $\mpl=1$ from now on)
\be
\label{soln}
ds^2=-dt^2+a^2(t)\rmd \x^2,\quad \phi^0=\bar c\int\frac{dt}{a(t)},\quad\phi^i=x^i,\quad \sigma=\ln (ca),
\ee
where 
\be
\label{cbar}
\bar c=const=1 + \frac{\om}{\kappa }~ \frac{H^2}{m^2},
\ee
while $c$ is given in \eqref{c's} (subject to the conditions \eqref{cond2} for the dS space).

At the present intermediate stage, it will be more convenient to work in terms of conformal time $\tau$, transforming to an "almost unitary" gauge in which the background metric is $g_{\mn}=a^2(\tau)\eta_{\mn}$ and the auxiliary scalars are \emph{frozen} to their background values $\phi^0=\bar c\tau$, $\phi^i=x^i$. We define the perturbations of the dynamical fields in this gauge as follows,
\be
g_{\mn}=a^2(\eta_{\mn}+h_{\mn}), \qquad \sigma=\ln(ca)+\zeta.
\ee
The tensor $\tK$, up to quadratic order in perturbations is given by the following expression
\begin{align}
\tK^\mu_{~\nu}&=\delta^\mu_\nu-c~a~ e^{\zeta}\sqrt{\frac{1}{a^2}(\eta^{\mu\lambda}-h^{\mu\lambda}+h^{\mu\rho}h_{\rho}^{~\lambda}+\dots) \Sigma_{\lambda\nu}}\nn\\
&=\delta^\mu_\nu-c~ (1+\zeta+\frac{1}{2}\zeta^2+\dots)\sqrt{\Sigma^\mu_{~\nu}-h^{\mu\lambda}\Sigma_{\lambda\nu}+h^{\mu\rho}h_{\rho}^{~\lambda}\Sigma_{\lambda\nu}+\dots) }
\end{align}
where all indices are assumed to be raised/lowered with the flat Minkowski metric and $\Sigma^{\mu}_{~\nu}\equiv\p^\mu\phi^a\p_\nu\phi^b\eta_{ab}=diag(\bar c^2,1,1,1)$. We will also need an expansion to the quadratic order of the metric determinant
\be
\sqrt{-g}=1+\frac{1}{2}h+\frac{1}{8}h^2-\frac{1}{4}h^{\mn}h_{\mn}+\dots~.
\ee
The Einstein frame action we would like to perturb can be conveniently parametrized by separating the pure general relativity sector in the following way 
 \begin{align}
\label{sublambdatilde0}
S_E&= \frac{1}{2}\int \sqrt{-g}~  [ R-6 H^2] \nn \\
&+ \frac{1}{2}\int \sqrt{-g}\left[ 6 H^2 -\frac{\omega}{\mpl^2} g^{\mn}\p_\mu\sigma\p_\nu\sigma
- \frac{m^2}{4} \( \mathcal{U}_2(\tK) +\alpha_3 \mathcal{U}_3(\tK)+\alpha_4 \mathcal{U}_4(\tK) \) \right].
\end{align}
where $H$ denotes the Hubble parameter of a dS solution at hand. The first term describes pure GR on dS space (with a CC, consistent with the expansion rate), while the perturbations of the rest of the lagrangian will describe deviation from GR.
The quadratic perturbations of the second line of \eqref{sublambdatilde0} can be written as follows (note that the indices on metric perturbations are \emph{not} raised and prime denotes a derivative w.r.t. conformal time $\tau$)
\begin{align}
\label{pert_h}
S^{(2)}_E\supset \frac{1}{2}\int d^4x ~a^4\bigg\{  &\frac{\omega}{a^2} \(\zeta'^2-(\p_i \zeta)^2\) +\frac{\omega H}{a} (h_{00}+h_{ii})\zeta' - \frac{2\omega H}{a} h_{0i} \p_i \zeta +(\gamma_1 h_{00}+\gamma_2 h_{ii})\zeta\nn \\
&+\gamma_3 h_{00}^2+\gamma_4h_{00}h_{ii}+ \gamma_5 h_{0i}h_{0i}+\gamma_6 h_{ij}h_{ij}+\gamma_7 h_{ii}^2\bigg\},
\end{align}
With the definitions
\begin{align}
\beta_0&=4+3\alpha_3+4\alpha_4\nn\\
\beta_1&=2+3\alpha_3+6\alpha_4\nn\\
\beta_2&=\alpha_3+4\alpha_4\nn\\
\beta_3&=2+\alpha_3+\alpha_4 ~,\nn
\end{align}
the coefficients appearing in the two forms of the action are
\begin{align}
\gamma_1&=3m^2\kappa\nn\\
\gamma_2&=-\frac{1}{4} m^2 c\(9 \bar c c^2\beta_2-4 (1+2\bar c) c \beta_1+3 (2+\bar c)\beta_0\)\nn\\
\gamma_3&=\frac{\omega}{4}H^2\nn\\
\gamma_4&=-\frac{m^2}{2}\kappa\nn\\
\gamma_5&=\frac{\kappa}{1+\bar c} m^2 \nn\\
\gamma_6&=\frac{1}{16} m^2 \( 3 \bar c c^3 \beta_2 -2(1+3\bar c) c^2\beta_1 +3(3+2\bar c) c \beta_0-24 \beta_3\)-\frac{H^2}{4}(\omega+6)\nn\\
\gamma_7&=\frac{1}{16} m^2 \( 2 \bar c c^2\beta_1 -3 (1+\bar c) c \beta_0+12 \beta_3\)+\frac{H^2}{8}(\omega+6)\nn~.
\end{align}

\subsubsection*{The new decoupling limit}
For analyzing the constraints on the parameter space, coming from the requirement of the absence of ghosts on the dS vacua of QMG, we resort to the \stu treatment of the perturbations. To this end, it is useful to return to the cosmic coordinates and define the metric perturbation in the standard way, $g_{\mn}=g^{FRW}_{\mn}+h_{\mn}$, with $g^{FRW}_{\mn}$ denoting the standard background FRW metric, given in \eqref{soln}. The (canonically normalized) metric perturbation is decomposed as follows,
\be
\label{stu}
h_{\mn}=\bar h_{\mn}+\frac{\nabla_\mu A_\nu+\nabla_\nu A_\mu}{m},
\ee
where $\nabla$ is a covariant derivative w.r.t. the background metric and $A_\mu$ denotes a canonically normalized vector field, encoding information about the helicity-1 and helicity-0 modes of the graviton. Furthermore, the spatial part of the vector $A_\mu$ is decomposed into the irreducible representations of the spatial rotation group as follows,
\be
A_i=S_i+\p_i b~, \qquad \p_i S_i=0.
\ee
We will start by examining the perturbation spectrum in the high frequency, \textit{decoupling} limit, defined as\footnote{Note that this decoupling limit differs from the one considered on Minkowski space above. }
\be
\label{declim}
m\to 0, \qquad H\to 0, \qquad \frac{H}{m}\equiv u =\textit{fixed}~.
\ee
The GR part of the action is invariant under the substitution \eqref{stu}, while the scalar contribution to the perturbation lagrangian \eqref{pert_h}, after some rearrangements and using the explicit form for some of the $\gamma$-coefficients, is given in the limit \eqref{declim} as follows\footnote{In the decoupling limit \eqref{declim}, the scale factor can be replaced by unity up to the terms, suppressed in this limit.} 
\begin{align}
\label{lagrscal}
S^{dl}_E&\supset \frac{1}{2}\int d^4 x \bigg\{ \omega\(\dot\rho^2-(\p_i\rho)^2 \)+p (\p_i A_0)^2-2 q \dot A_0\Delta b+r (\p_0\p_i b)^2 \nn \\&+4(\gamma_6+\gamma_7)(\Delta b)^2    \bigg\}~,
\end{align}
where $\rho=\zeta+u A_0$ and the following notation for different coefficients has been used
\be
\label{constants}
p=\omega u^2+\frac{\kappa}{1+\bar c•}, \qquad q=\kappa \(1-\frac{1}{1+\bar c}\), \qquad r=\frac{\kappa}{1+\bar c}.
\ee
There are a few important observations one can make about the lagrangian for scalar perturbations \eqref{lagrscal}. First, we see that $\omega>0$ is required for avoiding a negative kinetic term for one of the scalars $\rho$. Moreover, another scalar $A_0$ is non-dynamical on the dS background! This is a clear manifestation (and a nice consistency check of the calculation) of the BD ghost-freedom of the theory - there can only be a single helicity-0 graviton and the fluctuations of the quasi-dilaton in the scalar sector propagating on \textit{any} background. Note that none of the scalars are coupled to the external stress tensor in the decoupling limit, meaning that in the full theory their coupling to external sources is strongly suppressed. This is a manifestation of the absence of vDVZ discountinuity. Last but not least, integrating out the auxiliary field $A_0$, one obtains for the kinetic term of the third scalar
\be
S^{dl}_E\supset \(r-\frac{q^2}{p}\) (\p_0\p_i b)^2+\dots
\ee
Using the explicit form of the constants \eqref{constants} as well as the relation \eqref{cbar}, one can check that, the coefficient in front of the $b$ - kinetic term vanishes! If the coefficient $\(\gamma_6+\gamma_7\)$ of the last term in \eqref{lagrscal} were nonzero, one would conclude that the single scalar degree of freedom $\rho$ remains propagating for $\omega \neq 0$ on any self-accelerating background (at least at the quadratic level). Interestingly enough however, one can check by using the background equations of motion that this coefficient vanishes\footnote{We thank Claudia de Rham  for pointing out that this has to be the case and Andrew Tolley for first showing that this indeed is the case.}, so that this particular decoupling limit has nothing to say about the dynamics of $b$; then this leaves an option of two scalar modes potentially propagating on the self-accelerated background. A more refined analysis, needed for uncovering this dynamics will be presented elsewhere\footnote{The loss of dynamics for various degrees of freedom on non-trivial backgrounds has been already encountered a number of times in the context of ghost-free massive gravity models \cite{ deRham:2010tw, Berezhiani:2011mt, Gumrukcuoglu:2011zh, Guido}, as well as ghosty extensions of the Fierz-Pauli model \cite{Tasinato:2012mf}. But as pointed out above, the present model, at least for a certain parameter space, may avoid this problem.}. 

As a last step, we remark on one more possible constraint on the parameter space of a consistent theory. It comes from demanding the absence of ghosts in the vector sector. It is straightforward to see (see also \cite{Blas:2009my}, for example) that the relevant condition is
\be
\gamma_5 \geq 0 .
\ee
One can check that it does not lead to any further constraints on top of \eqref{cond2}. Moreover, $\gamma_5=0$ translates into the following condition on the parameters (consistent with \eqref{cond2})
\be
\alpha_3<-4, \qquad \alpha_4=-1-\frac{3}{4}\alpha_3.
\ee
In this part of the parameter space, the kinetic term for the vector perturbations on the dS solution vanishes, leading to no propagating vector modes on the dS vacua of such theories.

\section{Conclusions and Future Directions}

We have presented an extension of ghost-free massive GR by a scalar field, dubbed here the quasi-dilaton, based on a new global symmetry \eqref{newglob} that appears in these theories. The symmetry largely determines the form of the action, including the couplings of matter to the rest of the fields. The theory enjoys all important properties of massive GR; it possesses a Hamiltonian constraint, responsible for the absence of the Boulware-Deser ghost; moreover, nonlinear dynamics of the helicity-0 graviton and the quasi-dilaton lead to the presence of the Vainshtein mechanism, screening extra forces associated with these fields at solar system distances. 

Most interestingly, and unlike massive GR, the theory does admit flat FRW cosmology with potentially interesting phenomenological implications. For a broad class of the parameter space,
\be
\alpha_3\neq 0, \qquad 0<\alpha_4<\frac{\alpha^2_3}{8}, \qquad 0\leq \omega< 6,
\ee
there exist selfaccelerated dS solutions with stable perturbations. Moreover, one combination of scalars becomes non-dynamical on these backgrounds (for a special choice of parameters, vector modes can also be made non-dynamical).

Similar to massive GR, the extra scalars do not couple to matter in the decoupling limit on self-accelerated backgrounds, implying a strong suppression for these couplings in the full theory (and the absence of the vDVZ discontinuity on such vacua). This leads to no easily observable fifth forces in the theory at hand.

Apart from the theoretical constraints, we anticipate a number of phenomenological ones as well, bounding different parameters of the theory. One interesting phenomenological implication of QMG for instance is a modified effective Newton's constant, governing the expansion of the universe (different from the effective Newton's constant for the solar system processes, or growth of fluctuations); this can be seen e.g. from the Friedmann equation \eqref{eq:frwc}. This would place mild constraints on the parameter $\omega$, coming from the Big Bang nucleosynthesis and CMB (essentially from the expansion rate at the corresponding times). 

We defer these and other phenomenological aspects of the model to a separate study.

\vskip 1cm
\noindent {\bf Acknowledgments:} We thank Claudia deRham and Andrew Tolley for useful discussions and suggestions. G. D'A. is supported by the James Arthur Fellowship.
GG is supported by NSF grant PHY-0758032 and the NASA grant NNX12AF86G S06. GG thanks the Physics Department of Columbia University and Yang Institute for Theoretical Physics at 
Stony Brook University for hospitality. LH is supported by the DOE and NASA under coorperative agreements DE-FG02-92-ER40699 and NNX10AH14G. D.P. is supported by the US Department of Energy under contract DOE-FG03-97ER40546, and would like to thank the Center for Cosmology and Particle Physics at New York University for hospotality.

\section*{Appendix A}

We show in this appendix that at the linear level, QMG propagates the five degrees of freedom of a massive graviton plus a massless quasi-dilaton on Minkowski space. Instead of adopting the canonical Hamiltonian approach, we will choose a quicker way, showing the presence of just enough number of constraints and the corresponding dispersion relations describing the desired degrees of freedom.

At the quadratic order, the Einstein frame Lagrangian \eqref{eq:einstein} reduces in the unitary gauge to the following expression,
\be
2\times \mathcal{L}_2=-\frac{1}{2} h^{\mn}(\mathcal{E} h)_{\mn}-\frac{m^2}{4}\((h_{\mn}-2\sigma\eta_{\mn})^2-(h^\mu_\mu-2\sigma\eta^\mu_\mu)^2\)
-\omega\p^\mu\sigma\p_\mu\sigma.
\ee
It is convenient for the present purposes to work with a (Jordan frame) redefined spin-2 field, $\bar h_{\mn}=h_{\mn}-2\sigma\eta_{\mn}$, in terms of which the Lagrangian rewrites as follows,
\be
2\times \mathcal{L}_2=-\frac{1}{2} \bar h^{\mn}(\mathcal{E} \bar h)_{\mn}+2\bar h^{\mn}(\p_\mu\p_\nu\sigma-\eta_{\mn}\Box\sigma)-\frac{m^2}{4}(\bar h_{\mn}^2-\bar h^2)
+(6-\omega)\p^\mu\sigma\p_\mu\sigma.
\ee
The equations of motion for $\bar h_{\mn}$ and $\sigma$ obtained by varying the latter Lagrangian are given respectively as follows (we do not distinguish between the upper and lower indices for notational simplicity),
\begin{align}
\Box \bar h_{\mn}-\p_\alpha\p_\mu \bar h_{\alpha\nu} -\p_\alpha\p_\nu \bar h_{\alpha\mu}+\eta_{\mn}\p_\alpha\p_\beta\bar h_{\alpha\beta}+\p_\mu\p_\nu\bar h-&\eta_{\mn}\Box\bar h-m^2(\bar h_{\mn}-\eta_{\mn}\bar h) \nn \\
+&4(\p_\mu\p_\nu\sigma-\eta_{\mn}\Box\sigma)=0,
\end{align}
\be
(12-2\omega)\Box\sigma=2(\p_\mu\p_\nu\bar h_{\mn}-\Box\bar h).
\ee
Taking the divergence of the first equation yields the constraint,
\be
\label{const1}
\p_\mu\bar h_{\mn}=\p_\nu\bar h,
\ee
while taking its trace it implies,
\be
\label{constr2}
\bar h =4\frac{\Box}{m^2}\sigma.
\ee
These constraints remove five degrees of freedom from the system, leaving 6 propagating modes. Using \eqref{const1} and \eqref{constr2} in the equations of motion, they can be reduced to the Klein-Gordon form,
\be
\Box\sigma=0,
\ee
\be
(\Box-m^2)\( \bar h_{\mn}-4\frac{\p_\mu\p_\nu}{m^2}\sigma \)=0~,
\ee
indicating that the spectrum consists of five massive and one massless degrees of freedom.

\end{document}